\documentclass[envcountsect]{llncs} 

\usepackage{amsfonts,amssymb}
\usepackage{stmaryrd}
\usepackage{amsmath}

\newcommand{\Set}{{\sf Set}}

\newcommand{\Rel}{{\sf Rel}}

\newcommand{\id}{{\rm id}}

\renewcommand{\Bbb}{\mathbb}

\newcommand{\CCc}{{\Bbb C}}
\newcommand{\DDd}{{\Bbb D}}

\newcommand{\NNn}{{\Bbb N}}

\newcommand{\RRr}{{\Bbb R}}

\newcounter{countroman}
{\begin{list}{{\rm (\roman{countroman})}}{\usecounter{countroman}}}%
{\end{list}}

\newcounter{countalpha}
{\begin{list}{(\alph{countalpha})}{\usecounter{countalpha}}}%
{\end{list}}

\newcounter{countalphabf}
{\protect\begin{list}{{\rm (}{\bf \protect\alph{countalphabf}}{\rm%
)}}{\protect\usecounter{countalphabf}}}% 
{\end{list}}

\mathcode`\<="4268 %left delimiter
\mathcode`\>="5269 %right delimiter
\mathchardef\gt="313E %relation >
\mathchardef\lt="313C %relation <
\newsavebox{\barr}
\savebox{\barr}{\hspace*{-9.5pt}\raisebox{1.25pt}{$\scriptscriptstyle%
|$}\hspace*{4.5pt}} 
\newsavebox{\barrleft}
\savebox{\barrleft}{\hspace*{-8.5pt}\raisebox{1.25pt}{$\scriptscriptstyle%
|$}\hspace*{10pt}}

\newcommand{\epi}{\to\hspace{-2ex}\rightarrow}

\newcommand{\ot}{\leftarrow}

 \def\pushright#1{{%              set up
    \parfillskip=0pt            % so \par doesnt push \square to left
    \widowpenalty=10000         % so we dont break the page before \square
    \displaywidowpenalty=10000  % ditto
    \finalhyphendemerits=0      % TeXbook exercise 14.32
    \leavevmode                 % \nobreak means lines not pages
    \unskip                     % remove previous space or glue
    \nobreak                    % don't break lines
    \hfil                       % ragged right if we spill over
    \penalty50                  % discouragement to do so
    \hskip.2em                  % ensure some space
    \null                       % anchor following \hfill
    \hfill                      % push \square to right
    {#1}                        % the end-of-proof mark (or whatever)
    \par}}                      % build paragraph

 \def\qed{\pushright{$\square$}\penalty-700 \smallskip}

\newenvironment{prf}[1]{\begin{trivlist} \item[{\bf ~Proof}#1.]}%
{\qed\end{trivlist}}

\newcommand{\beq}{\begin{equation}}
\newcommand{\eeq}{\end{equation}}
\newcommand{\ba}[1]{\begin{array}{#1}}
\newcommand{\ea}{\end{array}}
\newcommand{\bea}{\begin{eqnarray}}
\newcommand{\eea}{\end{eqnarray}}
\newcommand{\bear}{\begin{eqnarray*}}
\newcommand{\eear}{\end{eqnarray*}}
\newcommand{\bpr}{\begin{prf}{}}
\newcommand{\epr}{\end{prf}}
\newcommand{\bprf}[1]{\begin{prf}{#1}}
\newcommand{\eprf}{\end{prf}}

\newtheorem{thm}{Theorem}[section]
\newtheorem{cond}{}[thm]

\newtheorem{prenumb}[thm]{\hspace{-1ex}}

\newdimen\proofrulebreadth \proofrulebreadth=.05em
\newdimen\proofdotseparation \proofdotseparation=1.25ex
\newdimen\proofrulebaseline \proofrulebaseline=2ex
\newcount\proofdotnumber \proofdotnumber=3
\let\then\relax
\def\hfi{\hskip0pt plus.0001fil}
\mathchardef\squigto="3A3B
\newif\ifinsideprooftree\insideprooftreefalse
\newif\ifonleftofproofrule\onleftofproofrulefalse
\newif\ifproofdots\proofdotsfalse
\newif\ifdoubleproof\doubleprooffalse
\let\wereinproofbit\relax
\newdimen\shortenproofleft
\newdimen\shortenproofright
\newdimen\proofbelowshift
\newbox\proofabove
\newbox\proofbelow
\newbox\proofrulename
\def\shiftproofbelow{\let\next\relax\afterassignment\setshiftproofbelow\dimen0 }
\def\shiftproofbelowneg{\def\next{\multiply\dimen0 by-1 }%
\afterassignment\setshiftproofbelow\dimen0 }
\def\setshiftproofbelow{\next\proofbelowshift=\dimen0 }
\def\setproofrulebreadth{\proofrulebreadth}

\def\prooftree{% NESTED ZERO (\ifonleftofproofrule)
\ifnum  \lastpenalty=1
\then   \unpenalty
\else   \onleftofproofrulefalse
\fi
\ifonleftofproofrule
\else   \ifinsideprooftree
        \then   \hskip.5em plus1fil
        \fi
\fi
\bgroup% NESTED ONE (\proofbelow, \proofrulename, \proofabove,
\setbox\proofbelow=\hbox{}\setbox\proofrulename=\hbox{}%
\let\justifies\proofover\let\leadsto\proofoverdots\let\Justifies\proofoverdbl
\let\using\proofusing\let\[\prooftree
\ifinsideprooftree\let\]\endprooftree\fi
\proofdotsfalse\doubleprooffalse
\let\thickness\setproofrulebreadth
\let\shiftright\shiftproofbelow \let\shift\shiftproofbelow
\let\shiftleft\shiftproofbelowneg
\let\ifwasinsideprooftree\ifinsideprooftree
\insideprooftreetrue
\setbox\proofabove=\hbox\bgroup$\displaystyle % NESTED TWO
\let\wereinproofbit\prooftree
\shortenproofleft=0pt \shortenproofright=0pt \proofbelowshift=0pt
\onleftofproofruletrue\penalty1
}

\def\eproofbit{% NESTED TWO
\ifx    \wereinproofbit\prooftree
\then   \ifcase \lastpenalty
        \then   \shortenproofright=0pt  % 0: some other object, no indentation
        \or     \unpenalty\hfil         % 1: empty hypotheses, just glue
        \or     \unpenalty\unskip       % 2: just had a tree, remove glue
        \else   \shortenproofright=0pt  % eh?
        \fi
\fi
\global\dimen0=\shortenproofleft
\global\dimen1=\shortenproofright
\global\dimen2=\proofrulebreadth
\global\dimen3=\proofbelowshift
\global\dimen4=\proofdotseparation
\global\count255=\proofdotnumber
$\egroup  % NESTED ONE
\shortenproofleft=\dimen0
\shortenproofright=\dimen1
\proofrulebreadth=\dimen2
\proofbelowshift=\dimen3
\proofdotseparation=\dimen4
\proofdotnumber=\count255
}

\def\proofover{% NESTED TWO
\eproofbit % NESTED ONE
\setbox\proofbelow=\hbox\bgroup % NESTED TWO
\let\wereinproofbit\proofover
$\displaystyle
}%
\def\proofoverdbl{% NESTED TWO
\eproofbit % NESTED ONE
\doubleprooftrue
\setbox\proofbelow=\hbox\bgroup % NESTED TWO
\let\wereinproofbit\proofoverdbl
$\displaystyle
}%
\def\proofoverdots{% NESTED TWO
\eproofbit % NESTED ONE
\proofdotstrue
\setbox\proofbelow=\hbox\bgroup % NESTED TWO
\let\wereinproofbit\proofoverdots
$\displaystyle
}%
\def\proofusing{% NESTED TWO
\eproofbit % NESTED ONE
\setbox\proofrulename=\hbox\bgroup % NESTED TWO
\let\wereinproofbit\proofusing
\kern0.3em$
}

\def\endprooftree{% NESTED TWO
\eproofbit % NESTED ONE
  \dimen5 =0pt% spread of hypotheses
\dimen0=\wd\proofabove \advance\dimen0-\shortenproofleft
\advance\dimen0-\shortenproofright
\dimen1=.5\dimen0 \advance\dimen1-.5\wd\proofbelow
\dimen4=\dimen1
\advance\dimen1\proofbelowshift \advance\dimen4-\proofbelowshift
\ifdim  \dimen1<0pt
\then   \advance\shortenproofleft\dimen1
        \advance\dimen0-\dimen1
        \dimen1=0pt
        \ifdim  \shortenproofleft<0pt
        \then   \setbox\proofabove=\hbox{%
                        \kern-\shortenproofleft\unhbox\proofabove}%
                \shortenproofleft=0pt
        \fi
\fi
\ifdim  \dimen4<0pt
\then   \advance\shortenproofright\dimen4
        \advance\dimen0-\dimen4
        \dimen4=0pt
\fi
\ifdim  \shortenproofright<\wd\proofrulename
\then   \shortenproofright=\wd\proofrulename
\fi
\dimen2=\shortenproofleft \advance\dimen2 by\dimen1
\dimen3=\shortenproofright\advance\dimen3 by\dimen4
\ifproofdots
\then
        \dimen6=\shortenproofleft \advance\dimen6 .5\dimen0
        \setbox1=\vbox to\proofdotseparation{\vss\hbox{$\cdot$}\vss}%
        \setbox0=\hbox{%
                \advance\dimen6-.5\wd1
                \kern\dimen6
                $\vcenter to\proofdotnumber\proofdotseparation
                        {\leaders\box1\vfill}$%
                \unhbox\proofrulename}%
\else   \dimen6=\fontdimen22\the\textfont2 % height of maths axis
        \dimen7=\dimen6
        \advance\dimen6by.5\proofrulebreadth
        \advance\dimen7by-.5\proofrulebreadth
        \setbox0=\hbox{%
                \kern\shortenproofleft
                \ifdoubleproof
                \then   \hbox to\dimen0{%
                        $\mathsurround0pt\mathord=\mkern-6mu%
                        \cleaders\hbox{$\mkern-2mu=\mkern-2mu$}\hfill
                        \mkern-6mu\mathord=$}%
                \else   \vrule height\dimen6 depth-\dimen7 width\dimen0
                \fi
                \unhbox\proofrulename}%
        \ht0=\dimen6 \dp0=-\dimen7
\fi
\let\doll\relax
\ifwasinsideprooftree
\then   \let\VBOX\vbox
\else   \ifmmode\else$\let\doll=$\fi
        \let\VBOX\vcenter
\fi
\VBOX   {\baselineskip\proofrulebaseline \lineskip.2ex
        \expandafter\lineskiplimit\ifproofdots0ex\else-0.6ex\fi
        \hbox   spread\dimen5   {\hfi\unhbox\proofabove\hfi}%
        \hbox{\box0}%
        \hbox   {\kern\dimen2 \box\proofbelow}}\doll%
\global\dimen2=\dimen2
\global\dimen3=\dimen3
\egroup % NESTED ZERO
\ifonleftofproofrule
\then   \shortenproofleft=\dimen2
\fi
\shortenproofright=\dimen3
\onleftofproofrulefalse
\ifinsideprooftree
\then   \hskip.5em plus 1fil \penalty2
\fi
}

\usepackage{textcomp}

\newcommand{\pfn}[3]{\xymatrix@-.8pc{{#1}\ar@{->}[r]|-{|}^-{#2}&{#3}}}
\newcommand{\Pfn}[2]{\xymatrix@-1pc{{#1}\ar@{=>}[r]|-{|}&{#2}}}
\newcommand{\fn}[3]{\xymatrix@-.8pc{{#1}\ar@{->}[r]^-{#2}&{#3}}}
\newcommand{\bimatrix}[8]{\mbox{\begin{tabular}{|lr|lr|}
\hline
&$#2$ & & $#4$ \\
$#1$ & & $#3$ &\\
\hline
& $#6$ & & $#8$\\
$#5$ && $#7$ &\\
\hline
\end{tabular}}
}

\newcommand{\rel}[3]{\xymatrix@-.8pc{{#1}\ar[r]|-@{|}^-{#2}&{#3}}}
\newcommand{\rell}[3]{\xymatrix{{#1}\ar@{->}[r]|-{\scriptscriptstyle |}^-{#2}&{#3}}}
\renewcommand{\Rel}{{\sf FRel}}
\newcommand{\Drel}{{\sf SRel}}

\newcommand{\normalize}[1]{\left\lfloor #1 \right\rfloor}

\renewcommand{\to}{\xymatrix@C-.5pc{\ar[r]&}}
\renewcommand{\ot}{\xymatrix@C-.5pc{& \ar[l]}}
\newcommand{\tto}[1]{\xymatrix@C-.5pc{\ar[r]^-{#1}&}}
\newcommand{\oot}[1]{\xymatrix@C-.5pc{&\ar[l]_-{#1}}}
\newcommand{\mono}{\xymatrix@C-.5pc{\ar@{>->}[r]&}} 
\renewcommand{\epi}{\xymatrix@C-.5pc{\ar@{->>}[r]&}}
\newcommand{\mmono}[1]{\xymatrix@C-.5pc{\ar@{>->}[r]^-{#1}&}} 
\newcommand{\eepi}[1]{\xymatrix@C-.5pc{\ar@{->>}[r]^{#1}&}}
\renewcommand{\mapsto}{\xymatrix@C-.5pc{\ar@{|->}[r]&}}
\newcommand{\mmapsto}[1]{\xymatrix@C-.5pc{\ar@{|->}[r]^{#1}&}}
\newcommand{\inclusion}{\xymatrix@C-.5pc{\ar@{^{(}->}[r] &}}
\newcommand{\iinclusion}[1]{\xymatrix@C-.5pc{\ar@{^{(}->}[r]^{#1}&}}
\newcommand{\dtto}[2]{\xymatrix@C-.5pc{\ar@<.875mm>[r]^{#1} \ar@<-.875mm>[r]_{#2}&}}

\usepackage{xy} 
\xyoption{all} 
\CompileMatrices 

\begin{document} 
\title{A semantical approach\\ to equilibria and rationality}
\author{Dusko Pavlovic\thanks{Supported by ONR and EPSRC.}\\%
\institute{Kestrel Institute and
Oxford University}
\email{\small %Email:~
dusko\char64\{kestrel.edu,comlab.ox.ac.uk\}%
}}
\date{}
\maketitle

\bigskip
\begin{flushright}
\parbox{5.5cm}{{\it \footnotesize "An equilibrium does not appear because agents are rational, but rather agents appear to be rational because an equilibrium has been reached.[\ldots] The task for game theory is to formulate a notion of   rationality."}}\\[3ex] {\footnotesize Larry Samuelson
  \cite[p.~3]{SamuelsonL:EGES}} 
\end{flushright}

\begin{abstract}
Game theoretic equilibria are mathematical expressions of rationality. Rational agents are used to model not only humans and their software representatives, but also organisms, populations, species and genes, interacting with each other and with the environment. Rational behaviors are achieved not only through conscious reasoning, but also through spontaneous stabilization at equilibrium points.

Formal theories of rationality are usually guided by informal intuitions, which are acquired by observing some concrete economic, biological, or network processes. Treating such processes as instances of computation, we reconstruct and refine some basic notions of equilibrium and rationality from the some basic structures of computation. 

It is, of course, well known that equilibria arise as fixed points; the point is that semantics of computation of fixed points seems to be providing novel methods, algebraic and coalgebraic, for reasoning about them.
\end{abstract}

%\medskip

\section{Introduction}
Game theory studies distributed processes where the resources are shared among the agents with different, inconsistent, and often adversarial goals. Originally devised as a tool of economics, politics, and warfare, game theory recently became an indispensable tool of algorithmics, especially as the processes and the problems of computation spread from computers to networks. The other way around, the algorithmic aspects of game theory have attracted a lot of attention on their own, leading to fruitful interactions between economics and algorithmics \cite{BorodinA:Online,Nisan:AGT,TardosE:STOC04}.

In semantics of computation, often viewed as the stylistic dual of algorithmics, the paradigm of game also played a crucial role, and led to the solutions of some deep and long standing problems \cite{AJM,HO}; yet the resulting toolkit of {\em game semantics\/} \cite{AbramskyS:GS} remained largely disjoint from the game theoretic methods, and concerns. While this may very well be justified by the different, and perhaps even disjoint goals of game semantics and game theory, the growing importance of the computational aspects of game theory continues to spur the explorations of a different conceptual link: {\em If gaming is computation, which semantical and programming methodologies apply to it?} 

The present paper provides a belated account of some initial explorations in this direction, going back to a joint project with Samson Abramsky.  The upshot is that the basic models of computation readily extend to capture the basic notions of game theory: the tools for reasoning about {\em choice}, be it possibilistic or probabilistic, and the tools to compute fixed points of possibilistic and probabilistic processes, turn out to be readily applicable to designing and programming strategic behavior. The approach seems promising in both directions: on one hand, the semantical view of games provides a convenient formal framework for conceptual analyses and concrete computations of response  profiles and equilibria; the other way around, the game theoretic view of the computational processes opens an alley towards modeling a wide range of network interactions of increasing practical interest.

As a running example, we use what may be the smallest and the deepest problem of game theory: Prisoners' Dilemma. In its standard solution, traditional game theory recommends selfishness as the only rational strategy here, although "staying the course" of this strategy leads to an ostensive loss for everyone. Can our semantical tools dispel the irrationality of this standard solution, and provide a better model of rationality? We propose and analyze several refinements of the basic model of strategic reasoning, and show how the implementations of the optimization task of gaming can be refined, and their rationality improved. Some familiar semantical  tools turn out to allow computing more informative equilibria, e.g. where players' preferences are quantified, rather than just partially ordered, and where the payoffs can be used dynamically (e.g. reinvested, or discounted), and not just accrued. This seems to suggest that applying semantical methodologies to program strategies may offer some new solutions, besides being fun.

\subsubsection*{Outline of the paper.} Section \ref{Semantics} sketches a bird's eye view of program and process semantics, and points to the place of games in that landscape. In section \ref{Nondet}, we reconstruct the familiar notions of Nash equilibrium and evolutionary stable strategy, as they could be obtained by running relational (nondeterministic) programs. We also discuss some nonstandard equilibrium concepts, which can be easily designed and implemented in this framework. In section \ref{Random}, we lift these equilibrium concepts from the relational to a stochastic framework, where they can be obtained as stationary distributions of Markov chains. In order to remain close to the usual game-theoretic models, in both these sections games are viewed as stateless processes. In section \ref{Position}, we discuss the role of state, i.e. position, in semantics of gaming. Section \ref{Conclusions} summarizes the paper.

\section{Program and process semantics of games}\label{Semantics}
Semantics of a natural language evolves through speech and through use of the language. Semantics of a programming language requires moreover a design effort, because it concerns not only communication between people, but also programming computers, and they need to be designed before they are built. However, as the notion of a computer is changing from a machine in a box to a world wide network, the simple notion of a program diversifies. Some programs acquire strategic, i.e. game theoretic aspects. We sketch a way to capture these aspects in a well studied framework of fixed point semantics, where coalgebras are always present in one way or another.

\subsection{Program semantics}
In categorical semantics, a program is denoted by an arrow $\fn{A}{f}{B}$ in a category $\DDd$, where the objects $A$ and $B$ denote some data types, of the inputs and of the outputs of $f$, respectively. It is assumed that the category $\DDd$ has cartesian products , so that we can also represent a program $\fn{A\times C}{g}{B\times D\times X}$ with multiple inputs and multiple outputs. 

But running a program does not just map data to data; it also causes a whole range of other observable effects. E.g., a computation may not terminate, or it may terminate with several possible outputs for the same input; or it may change the state of the computer, or of another resource. Such  computational effects can be captured by computational {\em monads} \cite{MoggiE:notions}. Originally proposed as a tool of semantics, monads have been widely endorsed as a convenient programming tool \cite{Wadler:Monads}. In the meantime, an alternative presentation of essentially equivalent semantical structure has been proposed, in terms of {\em premonoidal categories\/} \cite{Power-Robinson}. The category $\DDd$ of data types, with cartesian products and simple {\em deterministic\/} maps between them, is extended to a category $\CCc$ with the same data types as objects, but with the computations with nontrivial effects as its morphisms. Along the inclusion $\DDd\inclusion \CCc$, the cartesian products of $\DDd$ are mapped into the {\em premonoidal\/} tensor products in $\CCc$. The relevant semantical structure is sometimes called {\em Freyd category} \cite{Power-Thielecke:Freyd}. 

While the models of games that involve some of the well-studied computational effects seem quite interesting for future research, in the present paper we only consider the simplest effects of the {\em choice\/} operations, and in particular of the possibilistic (relational) and probabilisitic (randomized) choice. The reason is that these choice operations already come about in game theory, so that we can display some familiar ideas from a slightly different angle.

From the rich tool chest of program semantics, we shall thus consider only two simple but fundamental categories of computations:
\begin{itemize}
\item $\Rel$ of finite sets and relations, and
\item $\Drel$ of finite sets and stochastic relations. 
\end{itemize}
A morphism in either of these categories will be denoted by a crossed arrow $\rel{}{}{}$. While a binary relation $\rel{A}{R}{B}$ in $\Rel$ can be viewed as a matrix $B\times A\tto{R}\{0,1\}$, a stochastic relation $\rel{A}{P}{B}$ in $\Drel$ is a matrix $B\times A\tto{P} [0,1]$, i.e. $P = (p_{ji})_{B\times A}$, where $p_{ji}\in [0,1]$ and
$
\sum_{j\in B} p_{ji} = 1
$ holds for all $i\in A$. Intuitively, the entry $p_{ji}$ can thus be viewed as the probability that the input $i\in A$ will result in the output $j\in B$. 
%The fact that these probabilities may not add up to 1 accounts for the fact that there may be no output. 
The composition in $\Drel$ is the matrix composition. Both categories of computations $\CCc = \Rel, \Drel$ have the same cartesian subcategory of deterministic maps $\DDd = \Set$. They both happen to be monoidal, rather than premonoidal.

\subsection{Processes and controls}
We model processes simply as programs that depend on a state, and may change it. If the state space is represented by an object $X$ in a category of computations $\CCc$, then a process is thus a morphism $\rel{A\times X}{R}{B\times X}$. Since every category of computations inherits along the inclusion $\DDd\inclusion \CCc$ the cartesian diagonals $\fn{A}{\delta}{A\times A}$ and the projections $\fn{A\times B}{\pi_1}{A}$ and $\fn{A\times B}{\pi_2}{B}$, we can separate the data part and the state part of a process as
\begin{itemize}
\item $R_B\ :\ \xymatrix{{A\times X}\ar@{->}[r]|-@{|}^-{R}&{B\times X}\ar@{->}[r]|-@{|}^-{\pi_1} & B}$ and
\item $R_X\ :\ \xymatrix{{A\times X}\ar@{->}[r]|-@{|}^-{R}&{B\times X}\ar@{->}[r]|-@{|}^-{\pi_2} & X}$.
\end{itemize}
A process can be ongoing, and its outputs may be used to determine the inputs to be fed back into it. This is expressed through {\em feedback\/} $\rel{B\times X}{\phi}{A}$. To stabilize a process, a {\em control\/} $\rel{X}{\gamma}{A}$ can be extracted as a fixed point 
\[
\prooftree
%\shiftleft 2.2em 
\prooftree
%\shiftright 1em 
\xymatrix@-.8pc{A \times X \ar@{->}[r]|-@{|}^-R & B\times X & & B\times X \ar@{->}[r]|-@{|}^-\phi & A}
\justifies
\xymatrix@-.8pc{A \times X \ar@{->}[r]|-@{|}^-R & B\times X \ar@{->}[r]|-@{|}^-\phi & A}
\endprooftree
\justifies
\xymatrix@-.8pc{X \ar@{->}[rr]|-@{|}^{\gamma = Fix_A(\phi\circ R)} \ar@{->}[d]|-@{|}_-{\left< \gamma,\id\right>}&& A\\
A\times X \ar@{->}[rr]|-@{|}_R &&B\times X\ar@{->}[u]|-@{|}_-\phi}
\endprooftree
\]
Such fixed point operations play a central role in modeling processes and controls. We shall see that they play a central role in modeling games. The fixed point operations in $\Rel$ and $\Drel$ are spelled out in the Appendix. 

For a categorical insider, we add that any Freyd-category \cite{Power-Thielecke:Freyd} with a family of Conway fixed point operators \cite{Bloom-Esik,Plotkin-Simpson} should suffice for the (as yet putative) research in {\em abstract game theory}. Equivalently, a traced Freyd category will do as well \cite{Benton-Hyland}.

\paragraph{\bf Examples.} The simplest example of a process is a {\em Mealy machine}. A deterministic one is simply a function $\fn{A\times X}{}{B\times X}$. A nondeterministic (possibilistic) one is a relation $\rel{A\times X}{}{B\times X}$. A probabilistic automaton can in principle be viewed as a stochastic relation of the same type, just in $\Drel$ rather than in $\Rel$. Other examples of processes include Markov chains\ldots and even games.

\subsection{Games as processes}\label{Games}
An $m$-player game is a process  $\rel{A\times X}{\varrho}{B\times X}$ where the inputs, the outputs and the state consist of $m$ components, i.e.
\[
A = \prod_{i\in m} A_i\qquad B = \prod_{i\in m} B_i \qquad X = \prod_{i\in m} X_i
\]
where we represent ordinals following von Neumann, in the form $m = \{0,1,\ldots, m-1\}$.
The inputs $A_i$ are thought of as the {\em moves\/} available to the $i$-th player; the outputs $B_i$ are her {\em payoffs}; the states in $X_i$ are the {\em positions\/} that she can observe. The payoff types $B_i$ are usually ordered, and this ordering expresses player's preference. 

Games can thus be viewed as a special case of controllable processes, described in the preceding section. The optimization task of control is, however, slightly different. First of all, it is distributed: instead of a {\em global control}, each player designs and implements an {\em individual strategy}. And secondly, these strategies are not designed using feedback, to respond to the outputs, but rather to respond to the inputs supplied by the other players:
\[
\prooftree
\prooftree
\prooftree
\xymatrix@-.8pc{A \times X \ar@{->}[r]|-@{|}^-\varrho & B\times X}
\justifies
\xymatrix@-.8pc{A_{-i} \times X_i \ar@{->}[rr]|-@{|}^-{RS_i} && A_i}
\using \mbox{ (\textborn)}
\endprooftree
\justifies
\xymatrix@-.8pc{A \times X \ar@{->}[rrrr]|-@{|}^-{RS = <RS_i\circ \pi_i>_{i\in m}} &&&& A}
\using \mbox{ (\textborn\textborn)}
\endprooftree
\justifies
\xymatrix@-.8pc{X \ar@{->}[rrrr]|-@{|}^-{RS^\bullet = Fix_A(RS)} \ar@{->}[drr]|-@{|}_-{\left<RS^\bullet,\id \right>}&&&& A\\
	&& A\times X \ar@{->}[urr]|-@{|}_-{RS}}
	\using \mbox{ (\textmusicalnote)}
	\endprooftree
\]
At step (\textborn), each player $i$ implements her rationality through a {\em response relation\/} $RS_i$ to all of the opponents' moves, which are chosen from 
\bear
A_{-i} & = &  \prod_{k\in m,\ k\neq i} A_k
\eear
At step (\textborn\textborn), the individual response relations $RS_i$ are gathered into the response profile $RS$, which is simply the $m$-tuple of the relations $RS_i$. 

Finally, at step (\textmusicalnote), the equilibrium $RS^\bullet$ is computed, as the fixed point of the response profile $RS$. The equilibrium is an $m$-tuple of relations $\rel{X}{RS^\bullet_i}{A_i}$. It tells, for each position from $X$, how will the game be played if each player responds by $RS_i$ to everyone else's responses $RS_{-i} = <RS_k\circ \pi_k>_{k\in m\setminus \{i\}}$. The equilibrium is thus  the global (social) result of the local (individual) preferences and of distributed reasoning (programming) in pursuit of these preferences.  

How does an equilibrium come about? The usual explanation is that each player $i$ knows everyone's preferences, and can thus construct $RS_k$, for all $k\in m$, on her own, and thus compute the profile $RS$ and the equilibrium $RS^\bullet = Fix(RS)$. But this explanation should be taken as a metaphor. In reality, equilibria are often reached e.g. in biological systems, and in other genuinely distributed processes, where the agents do not perform explicit local computations, or reason about each other. Moreover, even in the cases where all strategies and their fixed points could conceivably be computed at each node, the fact that there are usually many equilibria gives rise to the question how do the players coordinate to meet at one equilibrium. This is where game theory enters the conceptual realm of {\em "invisible hand"\/} and equilibrium selection  \cite{SamuelsonL:EGES}. Modeling such genuinely distributed processes is one of the most interesting challenges of the computational semantics of gaming.

In the rest of the paper, we explore more closely each of the steps in the above derivation of the response strategies and equilibria. We begin with step (\textborn), where each player "programs" her response to other players' possible moves.

%\paragraph{Remark.} In game theory, and in our model, all players move together, or strictly alternate. In game semantics, each player performs a number of moves, before passing control to the opponent. This is the essence of game semantics.

%One passes from one model to another by changing the notion of a move. E.g., it is customary in game theory to hide the dynamics of the game by representing an entire strategy as a single move. This is useful to bring a game to a normal form; but bad because it hides dynamics.

%A good modeling methodology should accommodate all choices.

\section{Strategies as nondeterministic programs}\label{Nondet}
To reconstruct the first concepts of standard game theory, we first consider one shot games, i.e. where $X = 1$. This is standard in traditional game theory, where it is assumed that each player chooses a strategy in advance, and plays it out no matter what. The notion of position, or state, is thus abstracted away. Another standard assumption is that the payoffs are uniquely determined for each player, by mapping each tuple of moves in $A$ to a tuple of payoffs in $B$. The game thus boils down to a function $\fn{A}{\varrho}{B}$. If there are 2 players, called 0 and 1, and if their payoffs are in $B_0 = B_1 = \RRr$, this gives the usual bimatrix form $\fn{A_0 \times A_1}{\varrho}{\RRr\times \RRr}$. If there are 3 players, the game can be viewed as 3 tri-matrices, etc.

Even if the game, represented by the functions that compute the payoffs, is completely deterministic, the fact that the players need to choose between the various possible moves makes their strategies into nondeterministic programs. In this section, we view them as relations; later we view them as stochastic relations.

\subsection{Designing and refining relational strategies}\label{resps}
We assume that the payoff types $B_i$ are ordered, and that the players prefer higher payoffs. Each of them thus programs a response strategy towards the goal of maximizing his payoffs. We first consider some simple implementations, and then show how they can be {\em refined}.

\paragraph{\bf A. Best response} simply maximizes the payoff
\begin{eqnarray*}
s_{-i}\ BR_i\ s_i & \iff & \forall t_i \in A_i.\ \varrho_i(t_i, s_{-i})\leq  \varrho_i(s_i, s_{-i})
\end{eqnarray*}

\paragraph{\bf B. Stable response} refines the view by taking the possible opponents' responses into account:
\begin{eqnarray*}
s_{-i}\ SR_i\ s_i & \iff & \forall t_i \in A_i.\ \varrho_i(t_i, s_{-i})\leq \varrho_i(s_i, s_{-i}) \wedge \\ && \left(\varrho_i(t_i, s_{-i}) = \varrho_i(s_i, s_{-i}) \Rightarrow\right.\\ 
&& \left.\forall t_{-i}\in A_{-i}.\ \varrho_i(t_i, t_{-i}) \leq
\varrho_i(s_i, t_{-i})\right)
\end{eqnarray*} 
The idea is that a stable response $s_i$ to $s_{-i}$ should remain optimal in some neighborhood of $s_{-i}$. When the payoff function $\varrho_i$ is linear, it is easy to prove that the above definition captures this. Indeed, if the opponents deviate from $s_{-i}$ and play $(1-\varepsilon) s_{-i} + \varepsilon t_{-i}$ for a small $\varepsilon\gt 0$, then a stable response $s_i$ will still be the best, because the validity of
\bear
(1-\varepsilon)\varrho_i (s_i, s_{-i}) + \varepsilon\varrho_i (s_i,
t_{-i}) & \geq & (1-\varepsilon)\varrho_i (t_i, s_{-i}) +
\varepsilon\varrho_i (t_i, t_{-i}) 
\eear
for all $t_i$, follows from the above definition of
$SR_i$.
%\footnote{This kind of reasoning originated Maynard-Smith's \cite{ESS} explorations of evolutionary games, albeit symmetric.} 

\paragraph{\bf C. Uniform  response} goes a step further by taking the opponents' {\em best\/} response into account:
\begin{eqnarray*}
s_{-i}\ UR_i\ s_i & \iff & s_{-i}\ BR_i\ s_i \wedge \\
 && \forall t_{-i}\in A_{-i}.\ s_i\ BR_{-i}\ t_{-i} \Rightarrow t_{-i}\ BR_i\ s_i
 \end{eqnarray*}
 where $s_i\ BR_{-i}\ t_{-i}$ abbreviates $\forall k\in m.\ k\neq i\Rightarrow (s_i,t_{-i,k})\ BR_k\ t_k$.  
The best response $s_i$ is thus required to remain optimal not only with respect to $s_{-i}$, but also with respect to the opponents' best responses to the profiles that include $s_i$. This is a rational, but very strong requirement: the relation $UR_i$ may be empty. We mention it as a first attempt to refine the response by anticipating opponents' responses to it. The next example proceeds in this direction, while assuring a nonempty set of responses.

 \paragraph{\bf D. Constructive response.}  While the uniform response captures the best responses to {\em all\/} of the opponent's responses, the constructive response relation also capture the responses that may not be the best responses to a fixed opponent's move, but are better than what the best response would turn into in
 the context of opponent's rational responses to it.
  
 \begin{eqnarray*}
s_{-i}\ CR_i\ s_i & \iff & \forall t_i\in A_i.\ \varrho_i (s_i,s_{-i}) \lt \varrho_i(t_i,s_{-i}) \Rightarrow\\
 && \exists t_{-i}\in A_{-i}.\ \varrho_{-i}(t_{i},t_{-i}) \gt \varrho_{-i}(t_{i},s_{-i}) \wedge \\ 
 & & \hspace{6em} \varrho_i(t_i,t_{-i}) \lt \varrho_i(s_i,s_{-i})
 \end{eqnarray*}

\paragraph{\bf Example.} Prisoners' Dilemma is a famous 2-player game, usually presented with a
single state $X=1$ and two moves, "$c$ooperate" and "$d$efect", thus $A_0 = A_1 = \{c,d\}$, and $B_0 = B_1 = \RRr$. Players' preferences are given by a payoff function  $\varrho = <\varrho_0,\varrho_1> : \{c,d\}\times \{c,d\} \to \RRr\times \RRr$ which can be presented by the bimatrix
\begin{center}
\bimatrix {10} {10} {0} {11} {11} {0} {1} {1}
\end{center} 
telling that $\varrho_0(c,c) = \varrho_1 = 10$, $\varrho_0(c,d) = 0$, $\varrho_1(c,d) = 11$ etc. The point is that players' local reasoning leads to globally suboptimal outcome: each player seems forced to play $d$, because he wins more whether the opponent plays $c$ or $d$; but if both players play $d$ they both win less than if they both play $c$. The constructive response allows the players to keep the strategy $c$
as a candidate solution.  Although $\varrho_0(c,c)$ gives lower payoff
than $\varrho_0(d,c)$, the player 0 knows that the profile $(d,c)$ is
unlikely to happen, because $\varrho_1(d,d)\gt
\varrho_1(d,c)$. So he keeps $(c,c)$ as better than $(d,d)$.
 
Of course, this form of rationality does not offer the worst case
protection, and may not seem rational at all, because there is no
guarantee that $(c,c)$ will happen either. Indeed,
$\varrho_{1}(c,d)\gt \varrho_{1}(c,c)$ is likely to motivate player 1
to defect, which leads to the worst outcome for player 0, since
$\varrho_{0}(c,d)\lt \varrho_0(c,c)$.
 
However, if player 1 follows the rationality of $CR$, and not $BR$,
then he'll also consider cooperating, because of the threat that
player 0 would retaliate in response to his defection, and
$\varrho_{1}(d,d)\lt \varrho_{1}(c,c)$. So the possibility of the
solution $(c,c)$ depends on whether the players share the same rationality $CR$. This sharing cannot be coordinated in the relational model of one-round Prisoners' Dilemma. We shall later see how some more precise models do allow this.

\subsection{Playing out the strategies: computing the equilibria}
For each of the described notions of response, we now consider the corresponding notion of equilibrium, derived at step (\textmusicalnote) in section \ref{Games}. The relational fixed point operators are described in the Appendix.

\paragraph{\bf A. Rationalizability and the Nash equilibrium.}
The Nash best response relations yield the system of $n$ relations
$BR$, which we write as

\begin{eqnarray*}
s\ BR\ t & \iff & \forall i\in n.\ s_{-i}\
BR_i\ t_i
\end{eqnarray*}
It is not hard to see that the strong fixed point (see appendix) yields the solutions of that system, i.e.
\begin{eqnarray*}
BR^\bullet s & \iff & s\ BR\ s\\
& \iff & \forall i\in n.\ s_{-i}\ BR_i\ s_i
\end{eqnarray*}
This, of course, means that $s$ is a {\em Nash equilibrium} \cite{NashJ:Eq}. On the other hand, the weak fixed point extracts the transitive closure
of $BR$, i.e. the smallest relation $BR^*$ satisfying 
\begin{eqnarray*}
BR^* s & \iff & \exists t.\ BR^* t \wedge
t BR s
\end{eqnarray*}
In game theory, the strategies $\{s_i\ |\ BR_i^* s_i\}$ are said to be {\em rationalizable} \cite{BernheimD:rationalizable,Pearce:rationalizable}. An $s_i$ is
rationalizable if and only if there is a rationalizable
counterstrategy $t_{-i}$ for which $s_i$ is the best response.

\paragraph{\bf B. Stable Strategies.}
Given
\begin{eqnarray*}
s\ SR\ t & \iff & \forall i\in n.\ s_{-i}\ SR_i\ t_i 
\end{eqnarray*}
the fixed point
\begin{eqnarray*}
SR^\bullet s & \iff & \forall i\in n.\ s_{-i}\ SR_i\ s_i \\
&\iff & \forall t\in A. \forall i\in n.\ \varrho_i(t_i, s_{-i})\leq\varrho_i(s_i,s_{-i}) \wedge\\
 &&\varrho_i(s_i, t_{-i}) =\varrho_i(s_i, s_{-i}) \Rightarrow\\
 &&  \forall t_{-i}\in A_{-i}.\ \varrho_i(t_i, t_{-i})\leq\varrho_i(s_i, t_{-i})
\end{eqnarray*}
is an {\em evolutionary stable strategy}, which is a straightforward generalization of the concept due to biologist John Maynard-Smith \cite{ESS}\footnote{He considered the symmetric case, where all players have the same preferences and the same choice of actions.}. On the other hand, the weak fixed point
\begin{eqnarray*}
SR^* s & \iff & \exists t. SR^* t \wedge t SR s
\end{eqnarray*}
yields the new class of {\em stably rationalizable
strategies}. Unfolding the above equivalence tells that $s$ is stably
rationalizable iff every $s_i$ is the best response for some stably
rationalizable $t_{-i}$, and {\em moreover}, whenever $t_i$ is another
best response to $t_{-i}$, as good as $s_i$, then $s_i$ is at least as
good as $t_i$ with respect to the other counterstrategies.

\paragraph{\bf C. Uniform equilibria and profiles.}
\begin{eqnarray*}
UR^\bullet\ (s_i,s_{-i}) & \iff & \forall i\in n.\ s_{-i}\ UR_i\ s_i\\
&\iff & \forall i\in n.\ s_{-i}\ BR_i\ s_i\ \wedge\\
&& \forall t_{-i}\in A_{-i}.\ s_i\ BR_{-i}\ t_{-i} \Rightarrow t_{-i}\ BR_i\ s_i
\end{eqnarray*}
where $BR_{-i}$ is like in \ref{resps}C.
A uniform equilibrium $s$ is thus a Nash equilibrium such that each its components $s_i$ is a uniform move, in the sense that it lies in the set 
\bear
U_i  & = & \{s_i \in A_i|\ \forall t_{-i}\in A_{-i}.\ s_i\ BR_{-i}\ t_{-i}\Rightarrow t_{-i} BR_i s_i\} 
\eear
A Nash equilibrium thus fails to be uniform whenever some opponent has an alternative best response. The uniformity of a response $i$-th player assures that it is the best response also with respect to such alternatives. In a sense, the uniformity requirement only eliminates the unreliable Nash equilibria from the search space.

The weak fixed point
\begin{eqnarray*}
UR^*\ s & \iff & \exists t. UR^* t \wedge t UR s
\end{eqnarray*}
yields the new class of {\em uniformly rationalizable
strategies}. Unfolding the above equivalence tells that $s$ is
uniformly rationalizable iff every $s_i$ is a uniform best response
for some uniformly rationalizable $t_{-i}$. 

\paragraph{\bf D. Constructive equilibrium.}
\begin{eqnarray*}
CR^\bullet\ s & \iff & \forall i. s_{-i} CR_i s_i
\end{eqnarray*}
As it stands, this equilibrium includes the Nash equilibria, {\em and\/} the fixed points of $CR$, chosen because they yield better payoff than the equilibria. While $CR$ itself does not guarantee the feasilibity of any $CR_i$-response of a particular player, the $CR$-equilibrium does guarantee that all players have the same $CR$-justification. 

\paragraph{\bf Remark.}
The above characterizations of equilibria guarantee provide no existence guarantees: e.g., the set $BR^\bullet$ of the Nash equilibria, of course, always exists, but it can be empty. The existence, of course, requires additional side conditions, such as the convexity of the set of strategies \cite{NashJ:Eq}.

\paragraph{\bf Example.} For Prisoners' Dilemma, both $(c,c)$ and $(d,d)$ are constructive equilibria. The former is unstable, since each player can improve her immediate payoff by defecting. This gain can be offset by the loss from retaliation, and can be irrational, especially if the value of $(c,c)$ is much larger than the value of $(d,d)$. 

But the relational view of the strategic choices cannot express these quantitative considerations. In the next section, we explore a refinement where they can be expressed.

\section{Strategies as randomized programs}\label{Random}
In this section, we consider the framework where the preferences are quantified: the strategic choices are expressed as probability distributions over the available moves. A strategy is thus a randomized program.\footnote{The payoff functions can also be viewed as randomized programs, capturing games that involve some form of {\em gambling}. But this leads to an essentially different type of game theory\cite{Dubins-Savage}.} --- Is it possible to improve the rationality of strategic behaviors by quantifying the preferences, and biasing them more towards the more favorable moves?

In the standard game theoretic reasoning, the payoffs are only used as a convenient way to express players' preference ordering. Indeed, any affine transformation of a payoff matrix represents the same game. In the present section, this is not the case any more. {\em We assume that all payoffs are non-negative}, and normalize them into probability distributions.

\paragraph{\bf  A. Best response distribution} is just a normalization of the payoff function:
\begin{eqnarray*}
s_{-i}\ BD_i\ s_i  & = & \frac{\varrho_i(s_i, s_{-i})}{\sum_{t_i\in A_i} \varrho_i(t_i,s_{-i})}
\end{eqnarray*}
where $\fn{A_i\times A_{-i} }{BD_i}{[0,1]}$ is viewed as a fuzzy relation $\rel{A_{-i}}{BD_i}{A_i}$. The idea is that $s_{-i}\ BD_i\ s_i$ (which can be viewed as the matrix entry in the row $s_i$ and the column $s_{-i}$) records not only that $s_i$ is the best response to $s_{-i}$, like $BR_i$ did, i.e. not just that $s_i$ is preferred to the other responses $t_i$; but $s_{-i}\ BD_i\ s_i$ also quantifies how much better is $s_i$ than $t_i$, in terms of the difference $(s_{-i}\ BD_i\ s_i)-(s_{-i}\ BD_i\ t_i)$.

\paragraph{\bf  B. Stable response distribution} measures not only how good is $s_i$ as a response to $s_{-i}$, but also how good is it, on the average, with respect to the other countermoves $t_{-i}$.
\begin{eqnarray*}
s_{-i}\ SD_i\ s_i  & = & \frac{\varrho_i(s_i, s_{-i})\cdot \sum_{t_{-i}\in A_{-i}} \varrho_i(s_i,t_{-i})}{\sum_{t_i\in A_i} \varrho_i(t_i,s_{-i}) \cdot \sum_{t_{-i}\in A_{-i}} \varrho_i(t_i,t_{-i})} 
\end{eqnarray*}
Like in the case of the relational stable response, if $s_i$ and $t_i$ yield are equally good as responses to $s_{-i}$, then $s_i$ remains a stable response if it is at least as good as $t_i$ with respect to all other countermoves $t_{-i}$. Moreover, $s_i$ will now remain stable even if it is not as good as $t_i$ with respect to each other $t_{-i}$, but just if it is as good {\em on the average}. In fact, if $s_i$ is much better than $t_i$ on the average, the probability $s_{-i}\ SD_i\ s_i$ may be greater than $s_{-i}\ SD_i\ t_i$, even if $\varrho_i(s_i,s_{-i})\lt \varrho_i(t_i,s_{-i})$.

\paragraph{\bf  C. Uniform  response distribution} multiplies the probability of a response $s_i$ to $s_{-i}$ by the payoffs from the response $s_i$ to all other countermoves $t_{-i}$, averaged by the likelihood that $t_{-i}$ may occur as the countermove against $s_i$, which is taken to be proportional with $\varrho_{-i}(s_i,t_{-i})$. 
\begin{eqnarray*}
s_{-i}\ UD_i\ s_i  & = & \frac{\varrho_i(s_i, s_{-i})\cdot \sum_{t_{-i}\in A_{-i}} \varrho_i(s_i,t_{-i}) \cdot \varrho_{-i}(s_i,t_{-i})}{\sum_{t_i\in A_i} \varrho_i(t_i,s_{-i}) \cdot \sum_{t_{-i}\in A_{-i}} \varrho_i(t_i,t_{-i})\cdot \varrho_{-i}(t_i,t_{-i})}
\end{eqnarray*}
If it happens that $s_i$ is a good response across all the best countermoves $t_{-i}$ against it, then $s_i$ is assigned a high uniform response probability. This was expressed in the uniform response relation too. The preference is now not only quantified, but also {\em smoothened out}, as to have a high uniform response probability as soon as is a good response to the likely countermoves just {\em on the average}.

\paragraph{\bf  D. Constructive response distribution}
To simplify {\bf notation} for a sequence of values $<f(s,y)>_{s\in A}$ renormalized into a probability distribution over $A$, we shall henceforth write
\bear
\normalize{f(s,y)}_s & = & \frac{f(s,y)}{\sum_{t\in A} f(t,y)}
\eear
The upshot is that we get $\sum_{s\in A} \normalize{f(s,y)}_s = 1$. The subscript $s$, denoting the renormalized variable, will be omitted when clear from the context. 

We also write
\bear
a_+ & = & \left\{\begin{array}{cl}
a & \text{ if } a\gt 0\\
0 & \text{ otherwise}
\end{array}\right\}\ = \ \frac{|a| + a}{2}
\eear

Now define
\begin{eqnarray*}
s_{-i}\ CD_i\ s_i  & = & \Bigg\lfloor \varrho_i(s_i, s_{-i})  \\
&& + \bigvee_{t_{i}\in A_{i}} \Big( \varrho_i(t_i,s_{-i}) - \varrho_{i}(s_i,s_{-i})\Big)_+ \cdot \\
&& \hspace{2em} \sum_{t_{-i}\in A_{-i}} \Big(\varrho_{-i}(t_i,t_{-i}) - \varrho_{-i}(t_i,s_{-i})\Big)_+ \cdot \\
&& \hspace{5.5em}\Big(\varrho_i(s_i,s_{-i}) - \varrho_{i}(t_i,t_{-i})\Big)_+
\Bigg\rfloor_{s_i}
\end{eqnarray*}
The idea behind constructive distribution is that the probabilistic
weight of $s_i$ as a response to $s_{-i}$ is now increased to equal the weight of a $t_i$ that may be a better response to $s_{-i}$ alone, but for which there is a threat of the countermoves $t_{-i}$, which are better for the opponent than $s_{-i}$, but worse for the player.

\paragraph{\bf Examples.} Response distributions for Prisoners' Dilemma are now
\begin{align*}
BD_i & = 
\begin{pmatrix}\frac{10}{21} & {0} \\[.75ex] {\frac{11}{21}} & 1 \end{pmatrix} 
& 
SD_i & = \begin{pmatrix}\frac{25}{58} & {0} \\[.75ex] {\frac{33}{58}} & 1 \end{pmatrix}\\[1ex]
UD_i & = \begin{pmatrix}\frac{1000}{1011} & {0} \\[.75ex] {\frac{11}{1011}} & 1 \end{pmatrix} & 
CD_i & = \begin{pmatrix}\frac{19}{30} & {0} \\[.75ex] {\frac{11}{30}} & 1 \end{pmatrix}
\end{align*}
where the columns represent the opponent's moves $c$ and $d$, the rows the player's own responses, and the entries the suggested probability for each response. 

\paragraph{\bf Stochastic equilibria.} Stochastic response profiles are Markov chains, and the induced equilibria are their stationary distributions. Playing out the randomized response strategies and computing stochastic equilibria is thus placed in a rich and well ploughed field \cite{NorrisJ:Markov}.

A stochastic Nash equilibrium is a uniform fixed point $BD^\bullet = Fix(BD)$, which can be computed as in Appendix B. Since each player participating in the profile $BD$ responds by a mixed strategy where the frequency of a move is proportional to the payoff that it yields, the condition $BD^\bullet = BD\circ BD^\bullet$ means that $BD^\bullet$ {\em maximizes everyone's average payoff}. Formally, this is a consequence of the fact that $BD$ is a stochastic matrix, and that 1 is its greatest eigenvalue, so that the images of the vectors in the eigenspace of 1 are of maximal length.

The stochastic equilibria $SB^\bullet$, $UB^\bullet$ and $CB^\bullet$ maximize players' average payoffs in a similar manner, albeit for more refined notions of averaging, captured by their more refined response distributions.

\paragraph{\bf Back to the Dilemma.} The strategies $UD$ and $CD$ above recommend cooperation as a better response to peer's cooperation. One might thus hope that, by taking into account the {\em average\/} payoffs, the stochastic approach may overcome the myopic rationality of {\em defection\/} as the equilibrium in Prisoners' Dilemma. Unfortunately, it is easy to see the only fixed point of any response disrribution in the form $RD = \begin{pmatrix}p & 0\\ 1-p & 1\end{pmatrix}$ is the vector  $\begin{pmatrix}0\\ 1\end{pmatrix}$, as soon as $p\lt 1$. Defection is the only equilibrium. 

Let us try to understand why. Suppose that it is assured that both players play a constructive strategy. If they both assume that the other one will cooperate, each of them will cooperate at the first step with a probability $\frac{19}{30}$, which seems favorable. However, under the same assumption, the probability that either of them will cooperate at both of the first {\em two\/} steps is $\left(\frac{19}{30}\right)^2 = \frac{361}{900}$, which is not so favorable. And it exponentially converges to 0. With a static strategy repeated over and over, any probability $p$ that the opponent will cooperate in one step leads to the probability $p^n$ that he will cooperate in $n$ steps, which becomes 0 in the long run.  Trust and cooperation require memory and adaptation, which can be implemented in position games.

\section{Position and memory}\label{Position}
The positions in a position game $\fn{A\times X}{\varrho}{B\times X}$ are recorded in the state space $X = \prod_{i\in m} X_i$, where the projection $X_i$ shows what is visible to the player $i$. In games of perfect information, all of $X$ is visible to all players. Even if the payoff function $\fn{A\times X}{\varrho_B}{B}$, the players can use the positions to adapt their strategies, and $\fn{A\times X}{\varrho_X}{X}$ should update the position as each move is made.

For instance, in Iterated Prisoners' Dilemma, each player chooses a sequence of moves $\sigma_i = <s_i^0, s_i^1,\ldots, s_i^n>$ and collects at each step the payoff $\varrho(s_0^\ell, s_1^\ell)$. But the moves $s_i^\ell$ can be chosen adaptively, taking into account the previous $\ell$ moves. These moves can be recorded as the position. E.g., set $X = \left(\{c,d\}\times \{c,d\}\right)^\ast$, and besides the payoff bimatrix $\fn{\{c,d\}^2}{\varrho_{B}}{\RRr^2}$, declare the position update $\fn{\{c,d\}^2}{\varrho_{X}}{(\{c,d\}^2)^\ast}$ to be the list function $\varrho_X(s, x) = s::x$. What are the rational strategies now?

Axelrod reports about the Iterated Prisoners' Dilemma tournaments in \cite{Axelrod}. E.g., one of the simplest and most successful strategies was {\em tit-for-tat}. It uses a rudimentary notion of position, recording just the last move: i.e., $X = \{c,d\}^2$, and $\fn{\{c,d\}^2}{\varrho_X}{\{c,d\}^2}$ is the identity function. The tit-for-tat strategy $\rel{X}{TT_i}{A_i}$  is simply to repeat opponent's last move:
\[
<x_0, x_1>\ TT_0\ x_1\qquad\qquad <x_0, x_1>\ TT_1\ x_0
\]
If both players stick with this strategy, then they will 
\begin{itemize}
\item either forever cooperate, or forever defect --- if they agree initially, 
\item or forever alternate --- if they initially disagree. 
\end{itemize}
Within an $n$-round iterated game, this is clearly not an equilibrium strategy, since each player can win that game by switching any $c$ to $d$. However, both players' total gains in the game will be higher if they cooperate. That is why a cooperative strategy may be rational when the cumulative gains within a tournament are taken into account, while it may not be a rational way to win a single party of the same game, or to assure a higher payoff from a single move. 

The upshot is that a game, viewed in strategic form, may thus lead to three completely different games, depending on whether the payoffs are recorded per move, or per $n$ rounds against the same opponent, or per tournament against many opponents playing different strategies. While the different situations arguably determine different rationalities, which can be captured by different normal forms, the process view of a game, with the various positions through which it may evolve, displays not only the semantical relations between the different views of the same game, but also a dynamical view of adaptive strategies.

As a final example, consider a version of Iterated Prisoners' Dilemma, where the positions $X = \RRr\times \RRr$ record the cumulative gains of both players. The cumulative payoff function and the position update function thus happen to be identical,  $\fn{\{c,d\}^2\times \RRr^2}{\varrho}{\RRr^2}$. To give the game a sense of the moment, let us assume that the gains are subject to a galloping inflation rate of 50\% per round, i.e. that the cumulative payoffs are given by the bimatrix
\begin{center}
\bimatrix {10+\frac{x_0}{2}} {10+\frac{x_1}{2}} {\frac{x_0}{2}}
{11+\frac{x_0}{2}} {11+\frac{x_0}{2}} {\frac{x_1}{2}}
{1+\frac{x_0}{2}} {1+\frac{x_1}{2}} 
\end{center} 
where $x= <x_0,x_1>\in \RRr^2$ is the position, i.e. the previous gains. Suppose that the player uses the position-sensitive form of the constructive rationality
\begin{eqnarray*}
(s_{-i},x)\ CD_i\ s_i  & = & \Bigg\lfloor \varrho_i(s_i, s_{-i}, x)  \\
&& + \bigvee_{t_{i}\in A_{i}} \Big( \varrho_i(t_i,s_{-i},x) - \varrho_{i}(s_i,s_{-i},x)\Big)_+ \cdot \\
&& \hspace{2em} \sum_{t_{-i}\in A_{-i}} \Big(\varrho_{-i}(t_i,t_{-i},\xi) - \varrho_{-i}(t_i,s_{-i},\xi)\Big)_+ \cdot \\
&& \hspace{5.5em}\Big(\varrho_i(s_i,s_{-i},\xi) - \varrho_{i}(t_i,t_{-i},\xi)\Big)_+
\Bigg\rfloor_{s_i}
\end{eqnarray*}
where $\xi = \varrho_X(t_i,s_{-i},x)$ is the position reached after the profile $(t_i,s_{-i})$ is played at the position $x$. The response distribution is then
\bear
CD_i(x) &=& \begin{pmatrix}\frac{38+x}{60+2x} & \frac{x}{2+2x} \\[1ex] {\frac{22+x}{60+2x}} & \frac{2+x}{2+2x} \end{pmatrix}
\eear
Since $x$ changes at each step, the profile $CD$ is not a Markov chain any more. Its fixed point is cumbersome to compute explicitly, although it converges fast numerically. In any case, it is intuitively clear that the high inflation rate motivates the players to cooperate. If they do, the cumulative payoff for each of them approaches $\$ 10\cdot\sum_{k=0}^\infty \frac{1}{2^k} = \$ 20$. If they both defect, their cumulative payoffs are $\$ 1\cdot \sum_{k=0}^\infty \frac{1}{2^k} = \$2$. If they begin to cooperate and accumulate $\$ x $ each, and then one defects, he will acquire an advantage of $\$ 11$ for that move. But after 10 further moves, with both players defecting, his advantage will reduce to about $1$ cent, and the
cumulative payoff for both players will again boil down to $\$2$. 

\section{Conclusions and future work}\label{Conclusions}
We explored the semantical approaches to gaming from three directions: through relational programming of strategies in section \ref{Nondet}, through quantifying preferences in terms of distributions (rather than preorders) in section \ref{Random}, and finally by taking into account the positions and the process aspects of gaming, in section \ref{Position}. 

The advantage of viewing strategies as programs is that they can be refined, together with the notion of rationality that they express. To illustrate this point, we discussed in section \ref{Nondet} some simple refinements of the standard equilibrium concepts. 

The advantage of viewing strategies as {\em randomized\/} programs is that the problem of equilibrium selection \cite{SamuelsonL:EGES} can be attacked by the Markov chain methods. Mixed strategies are, of course, commonly used in game theory. They assure the existence of Nash equilibria. The mixture is interpreted either as a mixed population of players, each playing a single strategy, or as the probability distribution with which the single player chooses a single strategy \cite{ESS}. However, when equilibria are computed as stationary distributions of Markov chains, the mixture provides additional information which can be used to coordinate equilibrium selection. The concrete methods to extract and use this information need to be worked out in future research.

The most interesting feature of the semantical view of games is the dynamics of gaming, as it evolves from position to position. This feature has only been touched upon in the the present paper. On one hand, it leads beyond the Markovian realm, and equilibria are harder to compute. But on the other hand, in practice, the important rational solutions are often attained through genuinely adaptive, position sensitive strategies. The toy example of Prisoners' Dilemma already shows that a widely studied science of rationality may miss even the basic forms of social rationality because of small technical shortcomings. Combining semantics of computation and game theory may help eliminate them.

\subsubsection{Acknowledgement.} Through years, I have benefited from many conversations with Samson Abramsky, on a wide range of ideas about games and semantics. 

\bibliographystyle{abbrv}
\bibliography{game}

\appendix

\section{Appendix: Fixed points in $\Rel$}
For a relation $\rel{A\times X}{R}{A}$ in the monoidal category $(\Rel,\times,1)$, the standard fixed point operator (induced by its simple trace structure) gives $\rel{X}{R^\bullet}{A}$, defined
\begin{eqnarray*}
x R^\bullet a & \iff  & (x,a) R a
\end{eqnarray*}
On the other hand, the order structure of $\Rel$ induces another fixed point operator, where $\rel{X}{R^*}{A}$ is defined as the smallest relation satisfying $<id,R^*>R = R^*$, i.e.
\bear
x R^* a & \iff & \exists c\in A.\ x R^* c \wedge (x,c) R a
\eear
For each $x$, the set $xR^* = \{a\ |\ xR^*a\}$ is just the image of the transitive closure of $\rel{A}{xR}{A}$. It can be defined inductively, as
\begin{eqnarray*}
x R^* a & \iff  & \forall n\in \NNn.\ x R^n a,\mbox{ where}\\
xR^0 a & \iff & \exists a'\in A.\  (x,a') R a\\
xR^{n+1} a & \iff & \exists c\in A.\ xR^n c \wedge (x,c) R a
\end{eqnarray*}
or in terms of the image $xR(C) = \{a\ |\ \exists c\in C.\ (x,c)Ra\}$ and
\bear
xR^* & = & \bigcap_{n=1}^\infty xR^n(A)
\eear
If the containment order on $\wp A$ represents information, so the singletons $\{a\}$ are maxima, and $A$ is the minimum, then the above intersection is the least upper bound, and $R^*$ is the least fixed point. Indeed, the containment $R^\bullet \subseteq R^*$ means that in the information order $R^\bullet \sqsupseteq R^*$.

\section{Appendix: Fixed points in $\Drel$}
A stochastic matrix $\rel{k\times m}{H}{k}$ can be viewed as an $m$-tuple of square stochastic matrices $\rel{k}{H_i}{k}$. By the Perron-Frobenius theorem, each $H_i$ has $1$ as the principal eigenvalue. This can also be directly derived from the fact that the rows of each $H_i - I$ must be linearly dependent, since the sum of the entries of each of its columns is 0. The fixed vectors of each $H_i$ thus lie in its eigenspace of 1. But this space may be of a high dimension. Which $m$-tuple of vectors is the {\em uniform\/}
fixed point of $H$ \cite{Crole-Pitts:Fixp,Plotkin-Simpson}?

The uniform fixed points arise from the trace operations \cite{HasegawaM:uniformity,Benton-Hyland}. Let $\rel{k\times m}{H^\ast}{k}$ be formed from the projectors $\rel{k}{H^\ast_i}{k}$ to the principal eigenspaces of $\rel{k}{H_i}{k}$. The uniform fixed point $\rel{m}{H^\bullet}{k}$ of $\rel{k\times m}{H}{k}$ can be obtained by tracing out $k$ in $\widehat{H} = <H^\ast,H^\ast> : \rel{k\times m}{}{k\times k}$, defined by
\bear
\widehat{h}_{<v,w><u,i>} & = & h^\ast_{v<u,i>}h^\ast_{w<u,i>}
\eear
The uniform fixed point is thus $H^\bullet = \left(h^\bullet_{ui}\right)_{k\times m}$ where
\[
h^\bullet_{ui}  =   \normalize{\sum_{w\in
    k} \widehat{h}_{<w,u><w,i>}} = \frac{\sum_{w\in k}
  h^\ast_{w<w,i>}h^\ast_{u<w,i>}}{\sum_{v,w\in k} h^\ast_{w<w,i>}h^\ast_{v<w,i>}}  
\]
To check that this is a fixed point of $H$, i.e. that $H <H^\bullet, I> = H^\bullet : \rel{m}{}{k}$, note that $\widetilde{H} = <H^\bullet, I> : \rel{m}{}{k\times m}$ is  
\[
\widetilde{h}_{uji}\  = \  \left\{ \begin{array}{ll} h^\bullet_{ui} & \text{if $j=i$}\\
0 & \text{otherwise}
\end{array}\right\} \ = \ \left\{ \begin{array}{ll} \frac{\sum_{w\in
      k} h^\ast_{w<w,i>}h^\ast_{u<w,i>}}{\sum_{v,w\in k} h^\ast_{w<w,i>}h^\ast_{v<w,i>}}
  & \text{if $j=i$}\\ 
0 & \text{otherwise}
\end{array}\right.
\] 
Now $H \widetilde{H} = H^\bullet : \rel{m}{}{k}$ is satisfied iff  
%\bear
%h^\bullet_{ui} & = & \sum_{<v,j>\in k\times m} h_{u<v,j>} \normalize{\sum_{w\in k} h^\ast_{w<w,i>} h^\ast_{u<w,i>}} \\
%& = & \normalize{\sum_{w\in k} h^\ast_{w<w,i>} \sum_{v\in k} h_{u<v,i>} h^\ast_{v<w,i>}}
%\eear
\[
h^\bullet_{ui}\  = \ \sum_{<v,j>\in k\times m} h_{u<v,j>} \normalize{\sum_{w\in k} h^\ast_{w<w,i>} h^\ast_{u<w,i>}} \ =\  \normalize{\sum_{w\in k} h^\ast_{w<w,i>} \sum_{v\in k} h_{u<v,i>} h^\ast_{v<w,i>}}
\]
holds for each $i\in m, u\in k$. But this is
valid because $\sum_{v\in k} h_{u<v,i>} h^\ast_{v<w,i>} = h^\ast_{v<w,i>}$, i.e. $H_i H^\ast_i = H^\ast_i$ holds by the definition of $H^\ast$.
\end{document}